\magnification=\magstep1
\def\para{\par\noindent}
\baselineskip 20pt
\centerline{{\bf The Two Dimensional XY Spin Glass with}}
\centerline{{\bf Ferromagnetic Next-Nearest-Neighbour Interactions}}\para
\vskip 0.25cm
\para S.~Jain and K.~J.~Hammarling,
\para School of Mathematics and Computing,\para
University of Derby,\para
Kedleston Road,\para
Derby DE22 1GB,\para
U.K.\para
\vskip 0.25cm
\para
\vskip 2.0cm
\para PACS numbers: 75.50.Lk, 75.10.Nr, 02.70.Lq, 64.60.Cn, 75.10.Hk 
\vskip 4.0cm
\para Submitted for publication in Physical Review E (Rapid Communications)
\para Date: 8 May 1998
\vfill\eject
\para ABSTRACT
\para The random-bond XY spin glass with ferromagnetic
next-nearest-neighbour interactions is studied on a square lattice
by Monte Carlo
simulations. We find strong evidence for a finite-temperature spin glass
transition at $T_c\approx 1.1$. We also give estimates for the spin glass
critical exponents for different values of the strength of the 
nearest-neighbour interaction.
 Our results are consistent with universal behaviour.
\vfill\eject
\para Over the past two decades there has been considerable
 interest in spin 
glasses. There now exists substantial numerical evidence that
 the spin glass
transition occurs at zero temperature in two dimensions (2d) for both
 Ising [1-5] and
vector [6-9] spin glasses with nearest-neighbour interactions.
\para Spin glass order is widely believed to occur at a finite temperature for
the Ising spin glass in three dimensions [3,4]. Hence, the lower critical 
dimension of the Ising spin glass is expected to be between 2 and 3 [10,11]. 
\para The 
two-dimensional Ising spin glass has been re-investigated in recent
 years [12,13].
Shirakura and Matsubara [12] have presented numerical evidence to
 suggest that
the asymmetric random bond ($+J$ and $-0.8J$) Ising spin glass in 2d has a 
finite-temperature spin glass phase transition. 

Further evidence for a non-zero
transition for the Ising spin glass in 2d came recently from a Monte Carlo
study by Lemke and Campbell [13] who modified the Edwards-Anderson model to
include ferromagnetic interactions between next-nearest-neighbours. Lemke and
Campbell [13] found that the 
modified 2d Edwards-Anderson Ising spin glass exhibits behaviour very
 similar to that
seen in conventional Ising spin glasses in 3d [4].
\para For vector spin glasses, such as XY spin glasses, the
 situation is somewhat
more controversial and complicated because of the presence of chirality.
 The numerical evidence [6-9]
 clearly points to a
 zero-temperature transition in the 
 nearest-neighbour XY spin glass in 2d with both Gaussian and random ($\pm J$)
bond distributions. The chiral-glass transition is
  also believed to occur at zero-temperature [8,14,15].
 However, it has been argued
that the chiral and phase variables decouple on long length scales.
As a consequence, the values of the chiral- and spin-glass
correlation-length exponents are significantly different [8,15]. 

Although most work would seem to point towards a
zero-temperature phase transition also in three dimensions [6,7,16],
 some existing
Monte Carlo data
can be fitted equally well assuming a finite temperature transition [7]. 

Very
recently, Maucourt and Grempel [17] have used a domain wall renormalisation
 group method to suggest that the lower critical dimension of the XY spin glass
model is very close to three, even possibly three itself. The results, however,
point to a finite-temperature chiral-glass
transition in 3d [17].  
\para Given the recent developments for Ising spin glasses, in this Rapid
Communication we re-visit
the XY spin glass in 2d. Here we concentrate on the spin glass transition.
 We shall present numerical evidence that including
next-nearest-neighbour ferromagnetic interactions in a random-bond XY spin 
glass in 2d induces a finite-temperature spin glass transition.
\para The Hamiltonian for the modified model [13] we study is given by
$$ {\it H} = - \sum_{<i,j>} J_{ij} {\bf S}_i.{\bf S}_j\ -
 \sum_{<<k,m>>} {\bf S}_k.{\bf S}_m, \eqno(1)$$
where ${\bf S}_i$ are planar
 spins on a square lattice of size $L^2\ (L\le 12)$ with
periodic boundary conditions. Whereas the first summation $<i,j>$ runs over all
nearest-neighbour pairs only, the second summation $<<k,m>>$ denotes sums over
all second nearest-neighbour pairs a distance $\sqrt 2$ apart. The
 interactions $J_{ij}$ are quenched
independent random variables chosen from the following binary distribution:
$$ P(J_{ij}) = {1\over 2} [\delta (J_{ij} - \lambda ) + \delta (J_{ij}
 + \lambda)].\eqno(2)$$
In our simulations we consider $0\le\lambda\le 1$. Clearly,
 for $\lambda = 0$ the
lattice decouples into two independent inter-penetrating square sub-lattices.
Each sub-lattice is a pure XY model which will undergo a Kosterlitz-Thouless
 [18] phase transition at $T\approx 0.89$ [19].
\para In our Monte Carlo simulations we study the dimensionless Binder parameter
$g_{SG}$ given by [4]
$$ g_{SG} = 3 - 2{{q^{(4)}_{SG}}\over {(q^{(2)}_{SG})^2}}, \eqno (3)$$
where $q^{(2)}_{SG}$, the spin glass order parameter,
and $q^{(4)}_{SG}$ are
defined by
$$ q^{(2)}_{SG} = {1\over N^2}\sum_{i,j} [<{\bf S}_i.{\bf S}_j>^2_T]_J
 \eqno (4)$$
and
$$ q^{(4)}_{SG} = {1\over N^4}\sum_{i,j,n,p}
 [<{\bf S}_i.{\bf S}_j {\bf S}_n.{\bf S}_p>^2_T]_J \eqno (5)$$
and here $<...>_T$ denotes a thermal average, $[...]_J$ indicates an average
over disorder and $N = L\times L$.
\para The spin glass susceptibility, $\chi_{SG}$,
 is related to $q^{(2)}_{SG}$ by
$$\chi_{SG} = N q^{(2)}_{SG}. \eqno (6)$$
We analyse our results according to standard finite size scaling [4]. Near a
transition temperature $T_c$ the Binder parameter is expected to scale as
$$ g_{SG}(L,T) = {\overline g}_{SG}(L^{1/\nu}(T - T_c)), \eqno (7)$$
where ${\overline g}_{SG}$ and $\nu$ are the scaling
 function and the correlation
length exponent, respectively. The scaling form of the Binder parameter can be
used to determine the value of $T_c$ as
 $g_{SG}(L,T_c) = {\overline g}_{SG}(0)$
is independent of the system size $L$.
\para The scaling form of $\chi_{SG}$ is given by
$$\chi_{SG}(L,T) = L^{2-\eta}{\overline\chi}_{SG} (L^{1/\nu}
 (T - T_c)), \eqno (8)$$
where ${\overline\chi}_{SG}$ is the scaling function and $\eta$ is the exponent
describing the power-law decay of correlations at $T_c$. The value of $\eta$
can be determined from
$$\chi_{SG}(L,T_c) = L^{2-\eta} {\overline\chi}_{SG} (0). \eqno (9)$$
\para We now turn to our computer simulations and discuss the results.
We use Metropolis dynamics and sequential updating in our Monte Carlo
simulations. The method of Bhatt and Young [4] is used to ensure that
thermal equilibrium is achieved in the simulations.

For each value of $L\ (4\le L\le 12)$ and $\lambda\ (0.0\le\lambda\le 1.0)$
 we averaged over 500 pairs of
samples. As most of the simulations were performed over a relatively high
temperature range $(0.7\le T\le 1.4)$, we did not encounter any serious 
 equilibration problems. In fact, we were able to achieve equilibrium within
6400 Monte Carlo steps for the largest lattice at the lowest temperature 
considered. Nevertheless, the simulations presented in this work took 
approximately 3 months of CPU time distributed over 10 Silicon Graphics
workstations.

Plots of the Binder parameter for various values of $\lambda$ are
 shown in figures 1(a)-(e). The statistical error-bars are in most cases smaller
than the size of the data points.
 Each figure shows the data against the temperature
for the four different values of $L\ (L = 4, 6, 8$ and $12)$.
 The curves for $\lambda = 1.0$ (figure 1(a)),
$\lambda = 0.7$ (figure 1(b)), $\lambda = 0.5$ (figure 1(c))
 and $\lambda = 0.3$
(figure 1(d)) clearly intersect and splay out below the intersection point.
This strongly suggests a spin glass transition.
 Both the transition temperature,
$T_c(\lambda)$, and the value of the Binder parameter, $g_{SG}(T_c)$,
would appear to depend only marginly, if at all, on $\lambda$.
 Although there is some uncertainty in the
 intersection points for all of the values of $\lambda$, our data
 are consistent with a universal (i.e. $\lambda$-independent) transition
temperature. This is to be contrasted with the non-universal behaviour
found by Lemke and Campbell [13] in the Ising version of the model.
The values of $T_c(\lambda)$ are summarised in Table 1 which also contains
our estimates for the critical exponents (see below).

The curves in figure 1(e), which shows the plot for $\lambda = 0.0$, coalesce
at around $T\approx 0.9$ and then remain together for all lower temperatures.
(For this particular value of $\lambda$ we went down to a lower temperature,
$T = 0.5$.)
This is, of course, what one expects to occur at the Kosterlitz-Thouless
 [18] transition.

\para We next extract the critical exponents from the data. For illustrative
purposes, we show the scaling plots for $\lambda = 0.5$ only. Assuming that
 $T_c(\lambda = 0.5) = 1.1$, in figure 2 we give a log-log plot of $\chi_{SG}$
against $L$. The slope of the straight line yields a rather large value
 of $\eta = 0.96\pm
0.05$. The values we obtained for $\eta$ for the other cases of $\lambda$ can
be found in Table 1. Although there appears to be a lot of scatter in the values
for $\eta(\lambda)$, given the uncertainties in $T_c(\lambda)$, 
our estimates are not incompatible with a universal value.  

By setting the critical temperature to $T_c(\lambda = 0.5) = 1.1$ and
 considering $\nu$ as an adjustable parameter, we estimate the
 correlation-length exponent to be $\nu(\lambda = 0.5)=0.75\pm 0.1$.
 The error quoted here
is just an estimate of the range of values for which the data scale well. In
figure 3 we show a scaling plot of the Binder parameter
 against $(T-1.1) L^{1/0.75}$. We note that this value of $\nu$
 is very similar to that found recently [20] for the {\it four-dimensional} XY
spin glass. Our estimates for the correlation-length exponent for the other
values of $\lambda$ are given in Table 1. Once again, our results point to
a universal value of $\nu$.

Finally, as a consistency check, in figure 4 we display for $\lambda = 0.5$ a
 scaling plot of $\chi_{SG}/L^{2-\eta}$ against $(T-T_c) L^{1/\nu}$ with
$T_c = 1.1, \eta = 0.96$ and $\nu = 0.75$. The large error-bars on this plot
are a consequence of both the statistical errors in $\chi_{SG}$ and the
uncertainty in the value of $\eta$. Nevertheless, all of the data appear to
 scale reasonably well.
\para To conclude, we have presented data from Monte Carlo simulations of 
a short-range XY spin glass in 2d with ferromagnetic next-nearest-neighbour
interactions. We find strong evidence for a finite-temperature transition.
We have used finite-size scaling to estimate the spin glass
transition temperature and the critical exponents as functions of the strength
of the nearest-neighbour interaction. Our results are consistent with
universal behaviour. Furthermore, the value of the critical temperature and
the correlation-length exponent are very similar to those found recently for
the four-dimensional XY spin glass. 
\para We would like to thank Matthew Birkin for looking after the Silicon
Graphics workstations. KJH would like to acknowledge the University of
 Derby for a Research 
Studentship.
\vfill\eject
\para Table 1 
\para Estimates of the critical temperature, $T_c$, and the critical exponents,
$\eta$ and $\nu$, for various different values of $\lambda$.
\vskip 0.5cm
\vbox{\tabskip=0pt \offinterlineskip
\def\tablerule{\noalign{\hrule}}
\halign to250pt{\strut#& \vrule#\tabskip=1em plus2em&
\hfil#& \vrule#& \hfil#\hfil& \vrule#&
\hfil#& \vrule#& \hfil#& \vrule#\tabskip=0pt\cr\tablerule
&&\hidewidth $\lambda$\hidewidth&&
\hidewidth $T_c$\hidewidth&&
\hidewidth $\eta$\hidewidth&&
\hidewidth $\nu$\hidewidth&\cr\tablerule
&& 0.3&& $1.10\pm 0.05$&& $0.85\pm 0.06$&& $0.85\pm 0.05$&\cr\tablerule
&& 0.5&& $1.10\pm 0.05$&& $0.96\pm 0.05$&& $0.75\pm 0.1$&\cr\tablerule
&& 0.7&& $1.05\pm 0.10$&& $1.08\pm 0.06$&& $0.75\pm 0.05$&\cr\tablerule
&& 1.0&& $0.90\pm 0.15$&& $0.83\pm 0.05$&& $0.90\pm 0.1$&\cr\tablerule
\noalign{\bigskip}}}
\vfill\eject
 {\centerline FIGURE CAPTIONS}
\para Figure 1(a) 
\para A plot of the Binder parameter, $g_{SG}$, against the
temperature for $L = 4, 6, 8$ and $12$ with $\lambda = 1.0$. The lines
are just guides to the eye.
\vskip 0.5cm
\para Figure 1(b)
\para The Binder parameter against the temperature for $\lambda = 0.7$. The
lines are just guides to the eye.
\vskip 0.5cm
\para Figure 1(c)
\para The Binder parameter against the temperature for $\lambda = 0.5$. The
lines are just guides to the eye.
\vskip 0.5cm 
\para Figure 1(d)
\para The Binder parameter against the temperature for $\lambda = 0.3$. The 
lines are just guides to the eye.
\vskip 0.5cm
\para Figure 1(e)
\para A plot of $g_{SG}$ against the temperature. The lines, which are just to
guide the eye, clearly meet and then remain together for lower temperatures.
\vskip 0.5cm
\para Figure 2
\para A log-log plot of the spin glass susceptibility against the linear
dimension of the lattice for $\lambda = 0.5$.
 The slope of the straight line gives us a value of
$\eta = 0.96\pm 0.05$.
\vskip 0.5cm
\para Figure 3
\para A scaling plot of $g_{SG}$ against $(T-T_c) L^{1/\nu}$ with $T_c = 1.1$
and $\nu = 0.75$ (here $\lambda = 0.5$).
\vskip 0.5cm
\para Figure 4
\para A scaling plot of $\chi_{SG}/L^{2-\eta}$ versus $(T-T_c) L^{1/\nu}$ for
$\lambda =0.5$ assuming that $T_c = 1.1, \nu = 0.75$ and $\eta = 0.96$. The
solid line is just to guide the eye.
\vfill\eject
\par REFERENCES
\item {[1]} I.Morgenstern and K.Binder, Phys. Rev. B{\bf 22}, 288 (1980)
\item {[2]} A.J.Bray and M.A.Moore, J. Phys. C{\bf 17}, L463 (1984)
\item {[3]} R.R.P.Singh and S.Chakravarty, Phys. Rev. Letters {\bf 57},
 245 (1986) 
\item {[4]} R.N.Bhatt and A.P.Young, Phys. Rev. B{\bf 37}, 5606 (1988)
\item {[5]} Y.Ozeki, J. Phys. Soc. Jpn {\bf 59}, 3531 (1990)
\item {[6]} B.W.Morris, S.G.Colborne, M.A.Moore, A.J.Bray and J.Canisius, 
\item {} J. Phys. C{\bf 19}, 1157 (1986)
\item {[7]} S.Jain and A.P.Young, J. Phys. C{\bf 19}, 3913 (1986)
\item {[8]} P.Ray and M.A.Moore, Phys. Rev. B{\bf 45}, 5361 (1992)
\item {[9]} W.L.McMillan, Phys. Rev. B{\bf 31}, 342 (1985)
\item {[10]} K.Binder and A.P.Young, Rev. Mod. Phys. {\bf 58}, 801 (1986)
\item {[11]} E.Marinari, G.Parisi and J.J.Ruiz-Lorenzo, \lq{\it Numerical
Simulations of Spin Glass Systems}\rq , p130 in \lq {\bf Spin Glasses
 and Random
Fields}\rq , edited by A.P.Young (World Scientific, Singapore, 1997)
\item {[12]} T.Shirakura and F.Matsubara, Phys. Rev. Letters {\bf 79},
 2887 (1997)
\item {[13]} N.Lemke and I.A.Campbell, Phys. Rev. Letters {\bf 76}, 4616 (1996)
\item {[14]} H.Kawamura and M.Tanemura, J. Phys. Soc. Japan {\bf 54},
 4479 (1985); J. Phys. Soc. Japan {\bf 55}, 1802 (1986); Phys. Rev. B{\bf 36},
7177 (1987); J. Phys. Soc. Japan {\bf 60} 608 (1991)
\item {[15]} H.S.Bokil and A.P.Young, J. Phys. A{\bf 29}, L89 (1996)
\item {[16]} H.Kawamura, Phys. Rev. B{\bf 51}, 12398 (1995)
\item {[17]} J.Maucourt and D.R.Grempbell, Phys. Rev. Letters {\bf 80}, 770
 (1998)
\item {[18]} J.M.Kosterlitz and D.J.Thouless, J. Phys. C{\bf 6}, 1181 (1973)
\item {[19]} H.Weber and P.Minnhagen, Phys. Rev. B{\bf 37}, 5986 (1988)
\item {[20]} S.Jain, J. Phys. A{\bf 29}, L385 (1996)
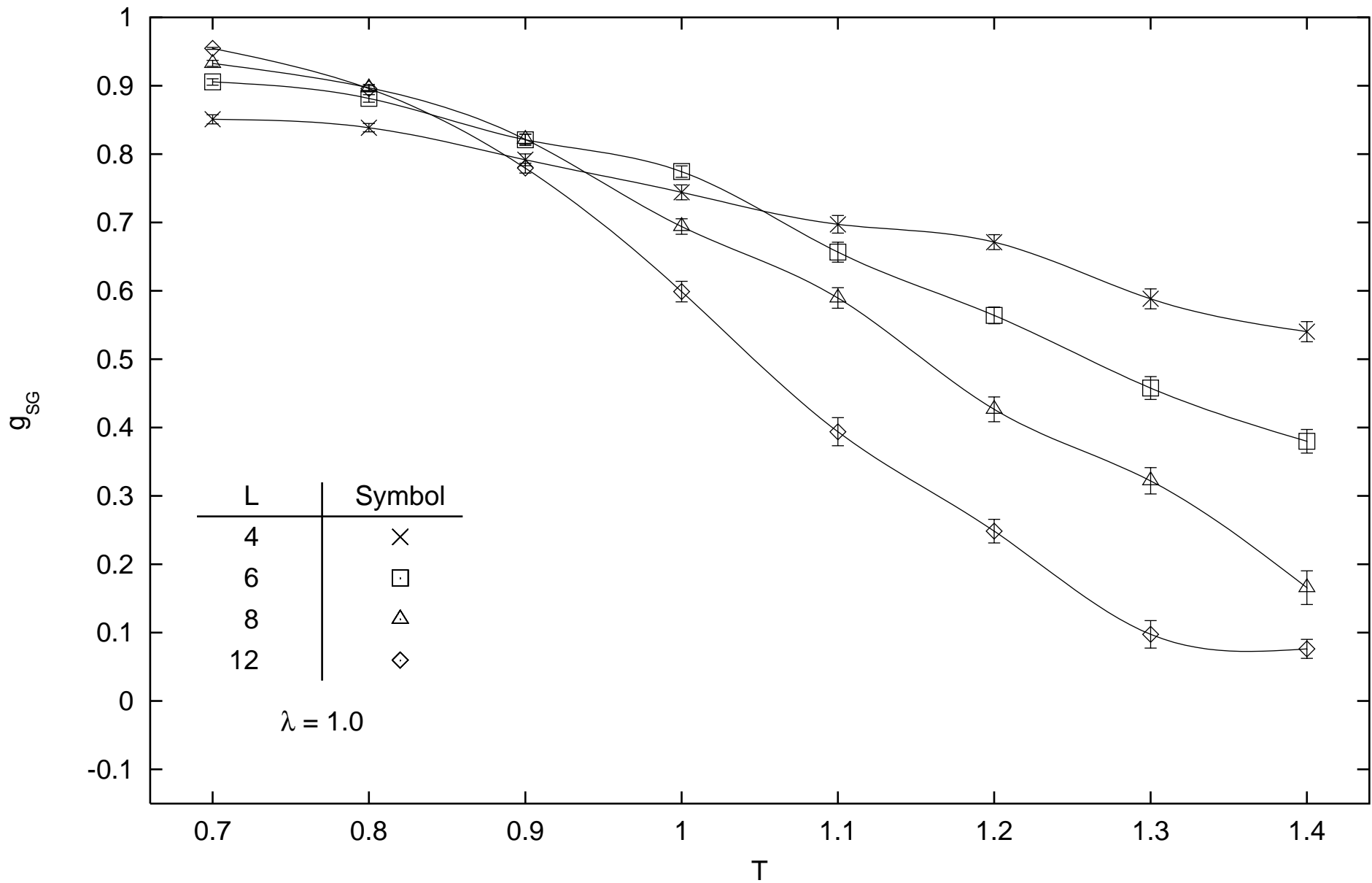
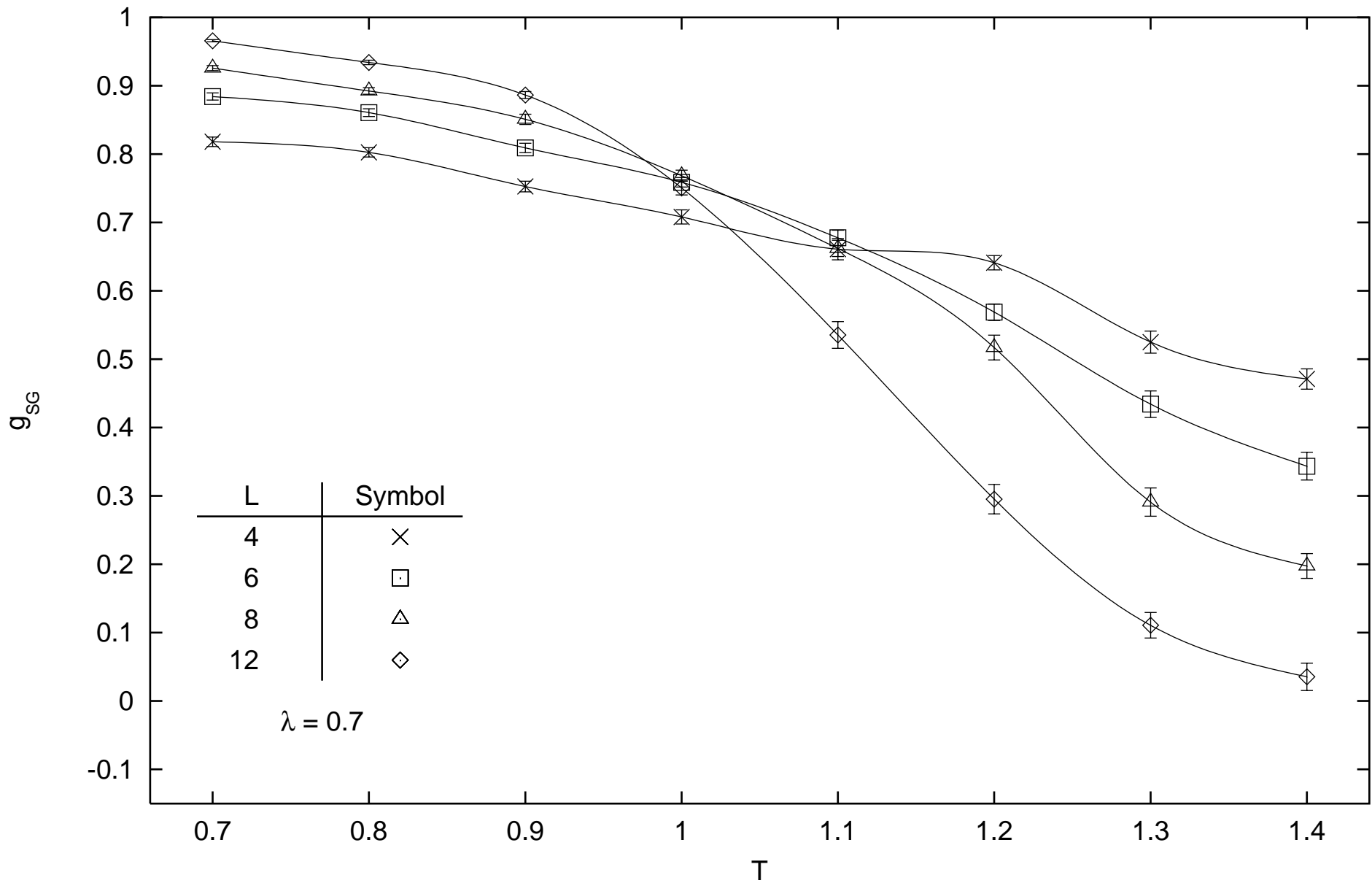
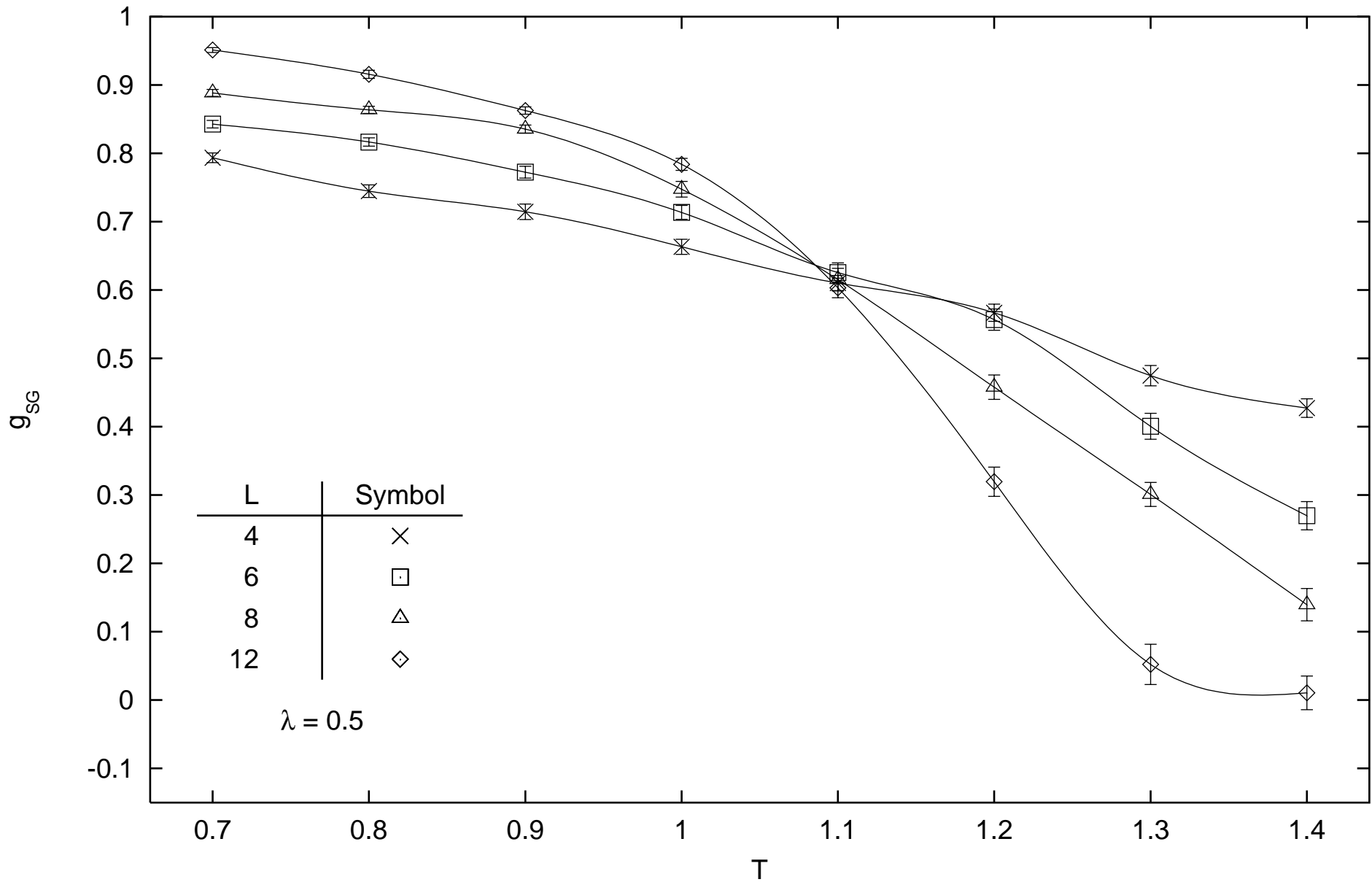
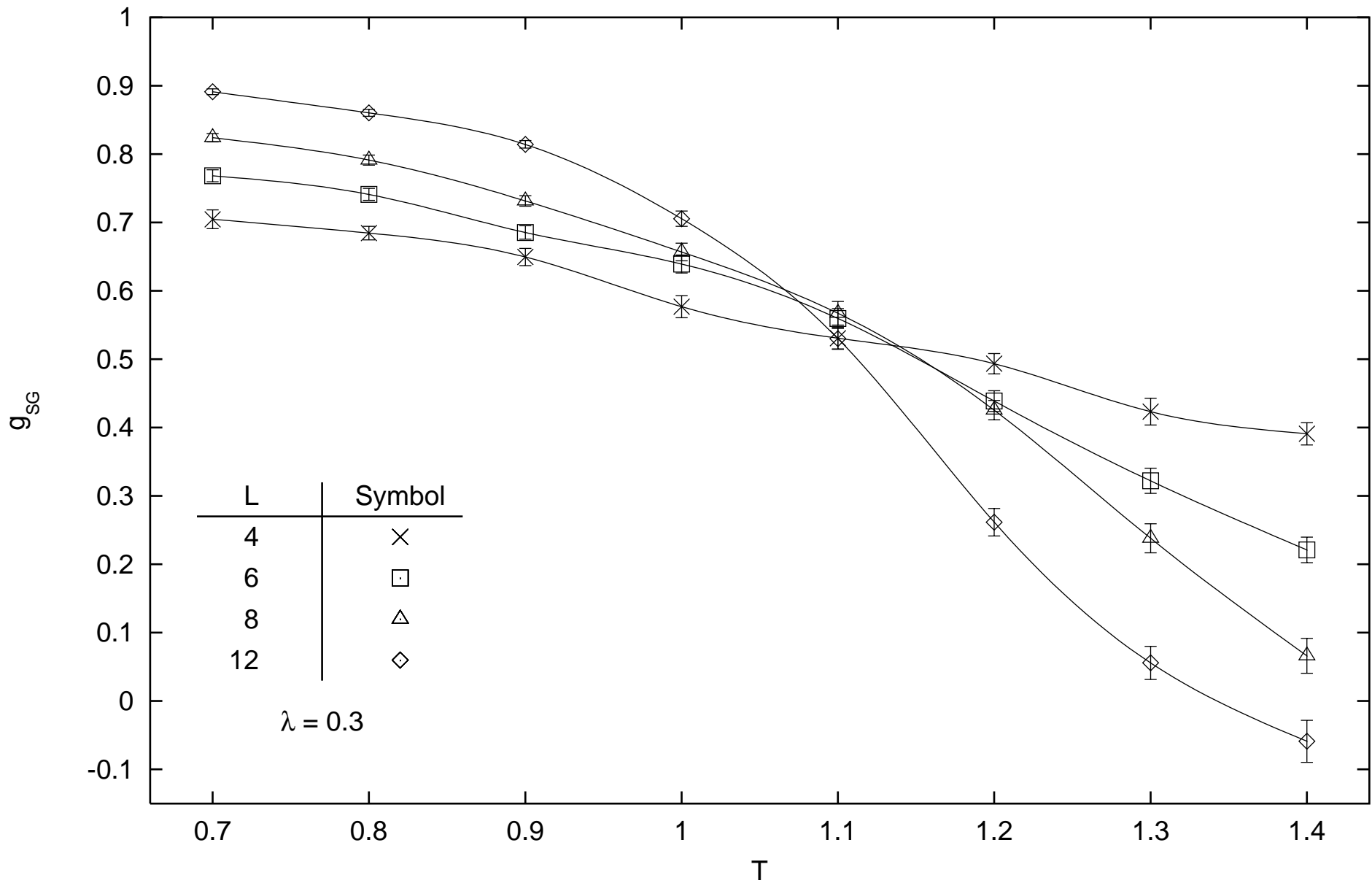
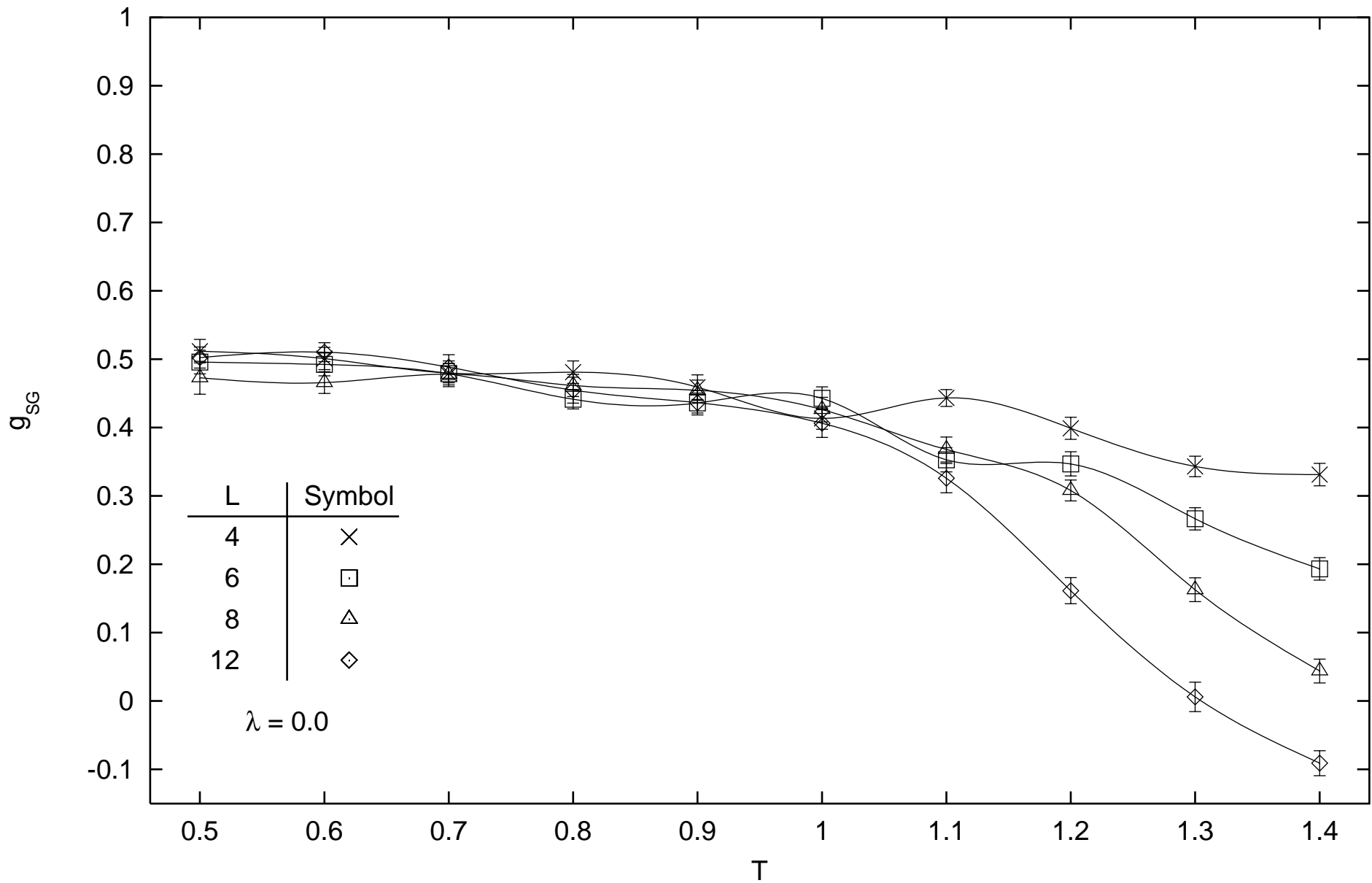
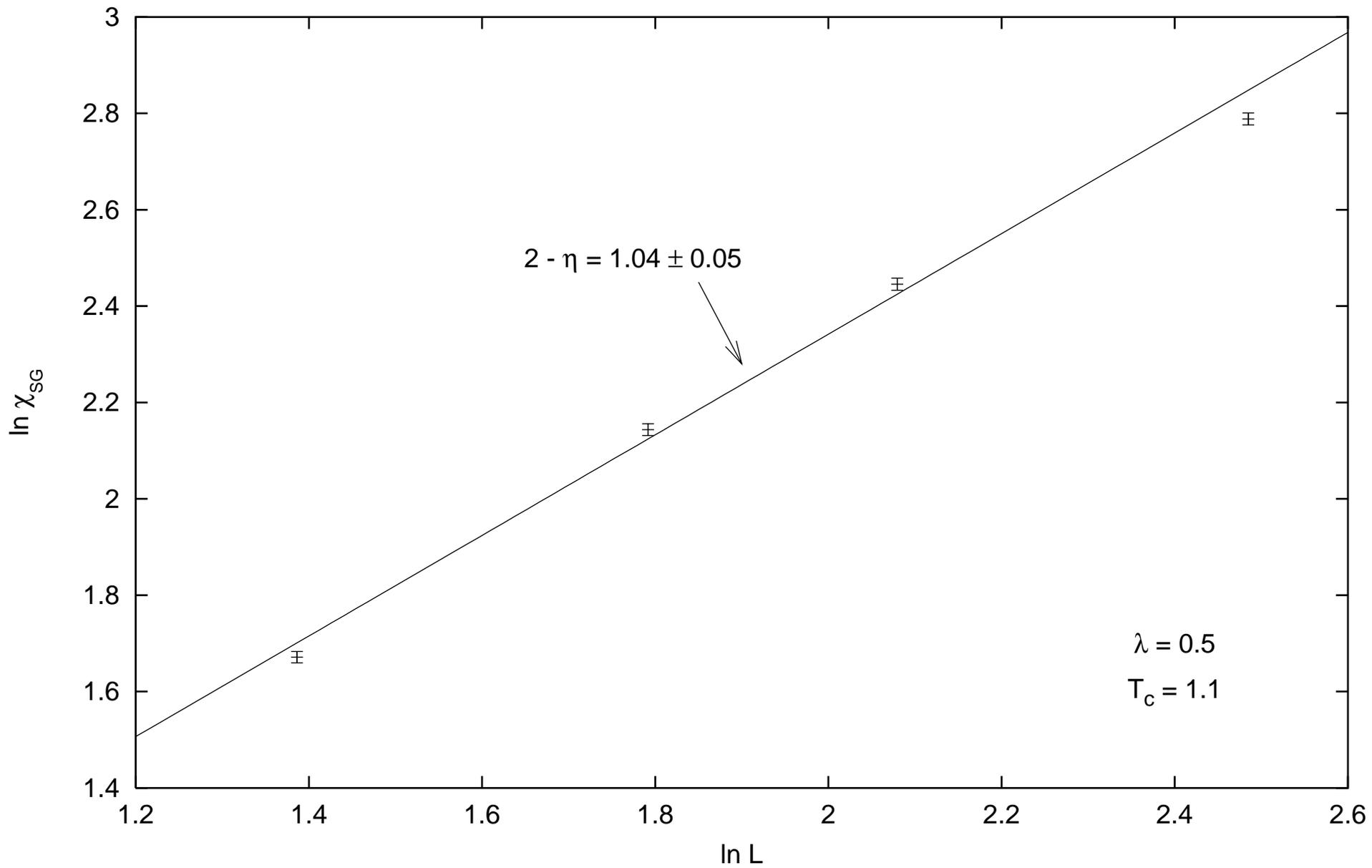
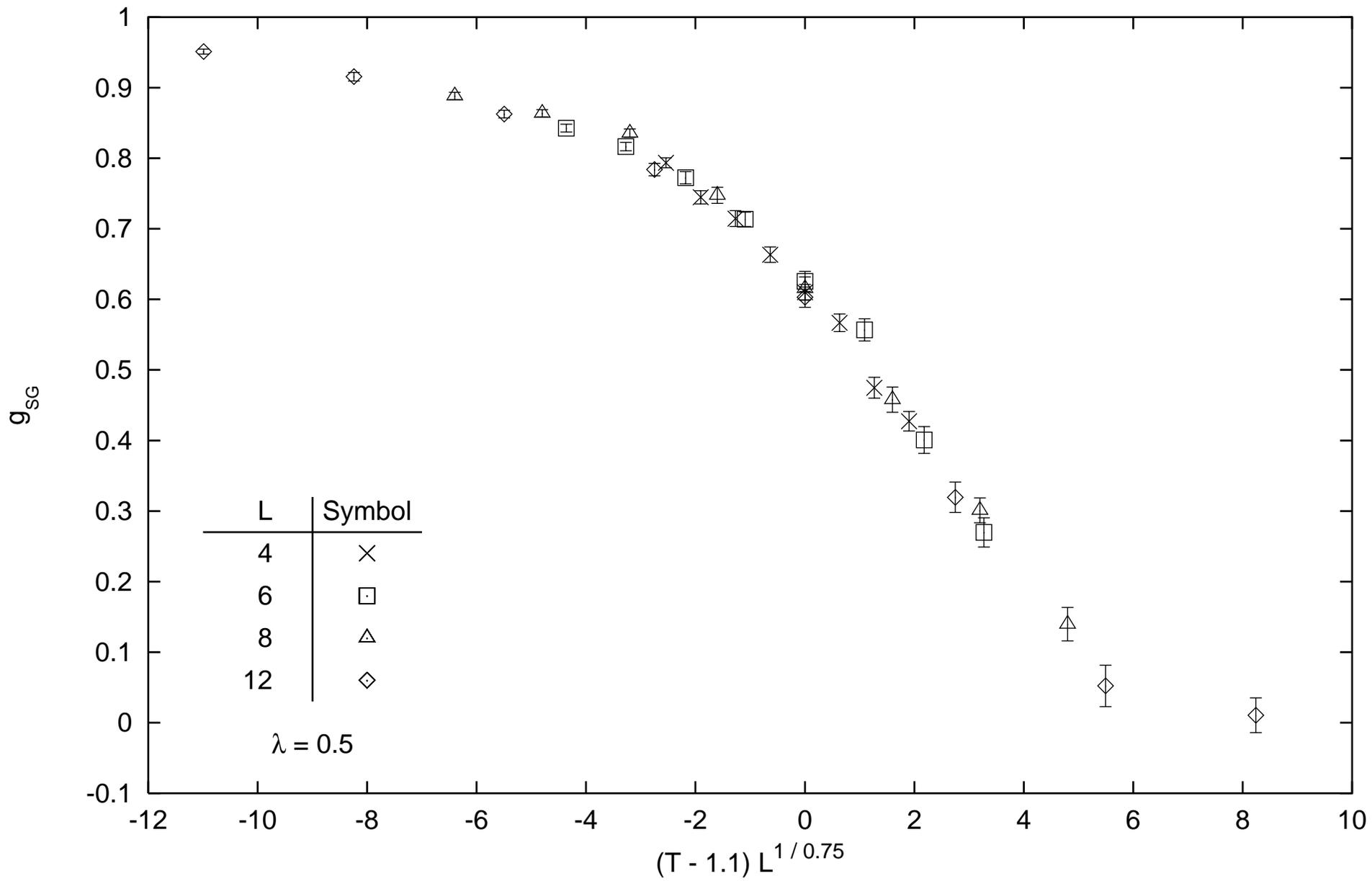
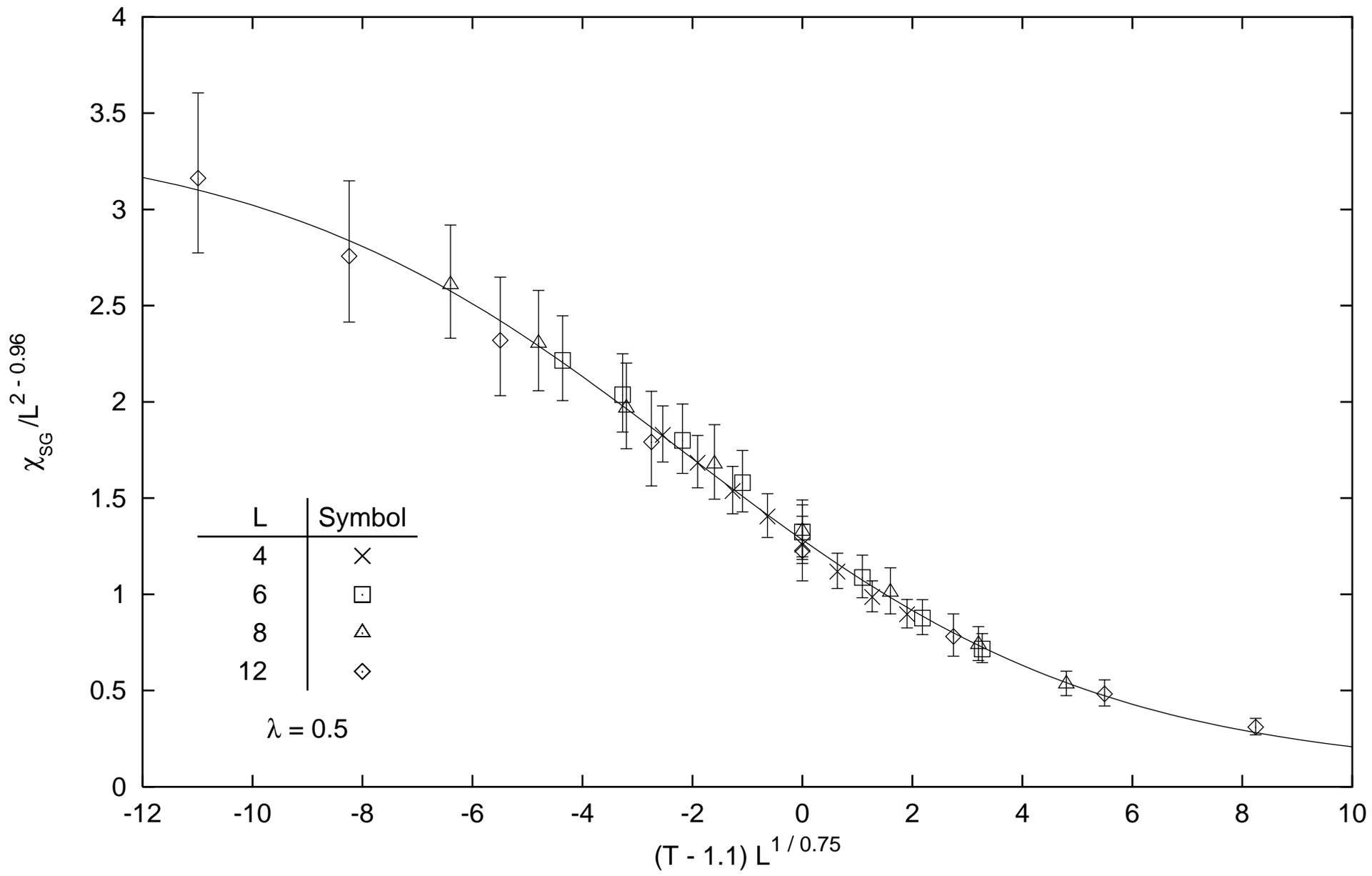
\end